# Ferroelectric properties of charge-ordered $\alpha$-(BEDT-TTF)$_2$I$_3$

P. Lunkenheimer[1,*], B. Hartmann[2], M. Lang[2], J. Müller[2], D. Schweitzer[3], S. Krohns[1], and A. Loidl[1]

[1]*Experimental Physics V, Center for Electronic Correlations and Magnetism, University of Augsburg, 86159 Augsburg, Germany*
[2]*Institute of Physics, Goethe-University Frankfurt, Max-von-Laue-Str. 1, 60438 Frankfurt(M), Germany*
[3]*Institute of Physics, University of Stuttgart, 70550 Stuttgart, Germany*

A detailed investigation of the out-of-plane electrical properties of charge-ordered $\alpha$-(BEDT-TTF)$_2$I$_3$ provides clear evidence for ferroelectricity. Similar to multiferroic $\kappa$-(BEDT-TTF)$_2$Cu[N(CN)$_2$]Cl, the polar order in this material is ascribed to the occurrence of bond- and site-centered charge order. Dielectric response typical for relaxor ferroelectricity is found deep in the charge-ordered state. We suggest an explanation in terms of the existence of polar and nonpolar stacks of the organic molecules in this material, preventing long-range ferroelectricity. The results are discussed in relation to the formation or absence of electronic polar order in related charge-transfer salts.

## I. INTRODUCTION

The polar order in electronic ferroelectrics arises from electronic degrees of freedom, in marked contrast to the off-center displacement of ions generating canonical ferroelectricity. In recent years, this exotic phenomenon has attracted considerable interest [1]. A promising route to electronic ferroelectricity is the combination of bond- and site-centered charge order (CO) [1]. Recent works have demonstrated that organic charge-transfer salts are good candidates for such a scenario [2,3,4,5,6,7,8,9,10]. A remarkable example is $\kappa$-(BEDT-TTF)$_2$Cu[N(CN)$_2$]Cl ($\kappa$-Cl), where BEDT-TTF stands for bis(ethylenedithio)-tetrathiafulvalene (often abbreviated as ET). In this material, the simultaneous occurrence of CO-driven ferroelectricity and magnetic order was reported, characterizing it as multiferroic [7]. However, although the simultaneous occurrence of ferroelectric and antiferromagnetic order has been unambiguously demonstrated, it should be noted that the presence of CO in $\kappa$-Cl is still controversially discussed [11,12,13].

In the present work, we provide dielectric and polarization data on $\alpha$-(ET)$_2$I$_3$, for which CO, leading to a pronounced metal-insulator transition below $T_{CO} \approx 135$ K [14,15], is a well-established fact [16,17]. Just as in $\kappa$-Cl, in $\alpha$-(ET)$_2$I$_3$ insulating anion sheets separate conducting layers formed by the ET molecules. The latter act as donors with an average charge of +0.5 per molecule. Within the planes, the ET molecules are arranged in a herringbone pattern with the long molecular axis oriented along the crystallographic $c$ direction, perpendicular to the planes, cf. inset of Fig. 1 [15]. The molecules form two types of alternating stacks oriented along the $a$ direction: Stack I is composed of weakly dimerized molecules denoted by A and A'. In contrast, the molecules in stack II (B and C) are not dimerized. While a weak charge disproportionation is already present at $T > T_{CO}$ [18], below the CO transition it becomes more pronounced.

For example, the charge values obtained from an X-ray study [17] are 0.82(9)$e$ (A), 0.29(9)$e$ (A'), 0.73(9)$e$ (B), and 0.26(9)$e$ (C). Obviously, CO occurs in both stacks but only in the dimerized stack I, well-pronounced ferroelectric-like order can be expected. However, it should be noted that the loss of inversion symmetry may also trigger small electronic or molecular deformations in the initially non dimerized stacks, and thus lead to the formation of weak polar order also along stacks II.

Based on optical second-harmonic generation (SHG) measurements, in Refs. [4] and [19], $\alpha$-(ET)$_2$I$_3$ indeed was shown to be a candidate for the occurrence of electronic ferroelectricity. SHG provides evidence for a non-centrosymmetric crystal structure, which is prerequisite for ferroelectricity [20,21]. However, while most of these structures are piezoelectric, ferroelectricity in addition requires a unique polar axis and the switchability of the polarization by an electrical field [22]. In Ref. [19], the domain boundaries revealed by SHG measurements were found to depend on the thermal history of the sample, already suggesting the controllability of the polarization. However, for a definite proof of ferroelectric ordering dielectric and polarization measurements are necessary. Interestingly, previous dielectric investigations of $\alpha$-(ET)$_2$I$_3$ did not reveal the typical signature of polar ordering, namely a peak in the temperature-dependent dielectric constant $\varepsilon'(T)$ [23,24]. Instead two relaxation modes with large amplitudes of dielectric constant and loss were detected and interpreted in terms of a "cooperative bond charge density wave with ferroelectriclike nature" [24]. These measurements were performed with the electric field oriented within the ET planes. Here, compared to typical ferroelectrics, the conductivity of $\alpha$-(ET)$_2$I$_3$ is high, even in the charge-ordered phase [24], making dielectric measurements a difficult task. For example, the high conductivity may obscure the signatures of ferroelectric order in the dielectric properties. Problems may also arise from non-intrinsic Maxwell-Wagner relaxations, which can lead to giant values of the dielectric constant [25,26]. Moreover, polarization measurements in conducting samples are strongly hampered by the shielding of the field arising from the mobile charge carriers.

Interestingly, the first dielectric measurements in $\kappa$-Cl were performed within the conducting ET planes, too, and quite similar large-amplitude relaxational behavior as in $\alpha$-$(ET)_2I_3$ was detected [27]. For $\kappa$-Cl, only dielectric measurements with the field directed perpendicular to the planes, where the conductivity is significantly reduced, were able to reveal the signatures of ferroelectric order [7]. This was further corroborated by nonlinear polarization measurements with out-of-plane field orientation [7]. It is clear that for both $\kappa$-Cl and $\alpha$-$(ET)_2I_3$, the ferroelectric polarization in principle should be mainly oriented parallel to the ET planes. However, as was noted in [7], in $\kappa$-Cl also an out-of-plane component of the polarization is expected, due to the inclined spatial orientation of the ET molecules [28]. A similar scenario may also apply for charge-ordered $\alpha$-$(ET)_2I_3$. Interestingly, in a recent work [29] it was shown that the hydrogen bonds connecting both ends of the A and A' molecules to the anion layers have different lengths. This leads to an asymmetric charge distribution along the long axis of the ET molecules, which is essentially oriented along **c**. Therefore, in the charge-ordered state the dipolar moments arising between the differently-charged molecules should also have a component perpendicular to the ET planes. Thus, it seems reasonable to perform dielectric and polarization measurements with the electrical field directed perpendicular to the ET planes, avoiding any problems arising from the high in-plane conductivity.

## II. EXPERIMENTAL DETAILS

Crystals of $\alpha$-$(ET)_2I_3$ were grown as reported in Ref. [15]. The geometry of the two investigated samples was plate-like (**c** axis vertical to the surface) with areas $A$ of about 3 and 2 mm$^2$ and thicknesses $d$ of about 30 and 50 μm for crystals 1 and 2, respectively. For the dielectric measurements, contacts of graphite paste were applied to opposite faces of the crystals, ensuring an electric field direction perpendicular to the ET-planes. The dielectric constant and conductivity were determined using a frequency-response analyzer (Novocontrol alpha-Analyzer). For the non-linear investigations, a ferroelectric analyzer (aixACCT TF2000) was used. Sample cooling was achieved by a $^4$He-bath cryostat (Cryovac) and a closed-cycle refrigerator.

## III. EXPERIMENTAL RESULTS

### A. Conductivity and permittivity

Figure 1 shows the temperature-dependent conductivity $\sigma'(T)$ measured at various frequencies. The results on both crystals agreed well (here those for crystal 1 are presented).

At temperatures above about 100 K, no significant frequency dependence of $\sigma'$ is detected. The CO phase transition is clearly revealed by a strong reduction of the conductivity below 133 K. At $T < T_{CO}$, $\sigma'(T)$ at the lowest frequency of 1 Hz (line in Fig. 1) shows an S-shaped decrease. Upon cooling, the conductivity curves at higher frequencies exhibit a frequency-dependent crossover to weaker temperature dependence where slight indications of shoulders, shifting with frequency, are found. Comparing the curves at the two lowest frequencies in Fig. 1 reveals that, at frequencies around 1 Hz, $\sigma'(T)$ is nearly independent of frequency. Thus, $\sigma'(T)$ at 1 Hz represents a good estimate of the dc conductivity $\sigma_{dc}(T)$. Until now, to our knowledge only the in-plane conductivity of $\alpha$-$(ET)_2I_3$ was reported (e.g., in [24,30]), which is several decades higher than the out-of-plane results presented here. As discussed in the Supplemental Material [31], we find that $\sigma_{dc}(T)$ for **E**||**c** does not follow thermally-activated Arrhenius behavior. Instead, variable-range-hopping conduction [32] may be valid at least for a limited temperature region.

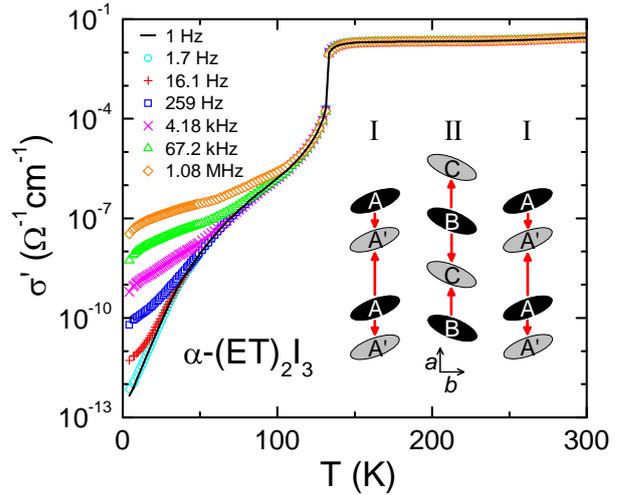

FIG. 1 (color online). Temperature dependence of the conductivity as measured for various frequencies. The inset provides a schematic representation of an ET plane (viewing direction along the long axis of the molecules). Black molecules have higher charge values. For clarity, the dimerization in stack I is strongly exaggerated; the variation in the relative orientations of adjacent molecules is not shown. The thick arrows indicate the dipolar moments between the molecules, adding up to a macroscopic polarization in the dimerized stacks.

Figure 2 shows the temperature-dependent dielectric constant of crystal 1 for various frequencies. Crystal 2 revealed very similar behavior [31]. Due to the rather high conductivity at $T > T_{CO}$, $\varepsilon'(T,\nu)$ could only be determined in the charge-ordered state. Depending on frequency and $T$, two very different relaxational behaviors can be observed. For $T \gtrsim 80$ K and frequencies $\nu \geq 22.1$ kHz, $\varepsilon'(T)$ exhibits a gradual steplike decrease upon cooling, shifting to higher temperatures with increasing frequency. This is the signature of relaxational behavior as typically arising from the reorientation of dipolar degrees of freedom [33,34]. Most interestingly, at lower frequencies a peak in $\varepsilon'(T)$ develops, which for the lowest frequencies is located at temperatures around 40 - 50 K. Its amplitude strongly increases and its position decreases with decreasing frequency. Moreover, the high-temperature flanks of all peaks share a common curve. The relation between peak temperature and frequency can be described by a Vogel-Fulcher-Tammann law [31,35], which is well established in glass physics [33,34] and also often employed for relaxor ferroelectrics [36,37,38]. Finally, the mentioned shoulders in the conductivity (Fig. 1) correspond to peaks in the dielectric loss which shift with frequency, indicating relaxational behavior [31]. All these findings are characteristic for so-called relaxor ferroelectrics, where the typical strong dispersion effects are usually ascribed to the freezing-in of short-range clusterlike ferroelectric order [39,40].

(a) and loss (b), and of the conductivity (c) for selected temperatures. $\varepsilon'(\nu)$ exhibits a gradual step-like decrease with increasing frequency, shifting to lower frequencies with decreasing temperatures. This is the typical signature of relaxational behavior in the frequency domain [33]. It is in full accord with the relaxational behavior seen in the temperature dependence of $\varepsilon'$ shown in Fig. 2. For the higher temperatures ($T \geq 79$ K), the low-frequency plateau of $\varepsilon'(\nu)$, corresponding to the static dielectric constant $\varepsilon_s$, only weakly varies with temperature, indicating conventional dipolar relaxation behavior. However, for lower temperatures $\varepsilon_s$ strongly varies with temperature, which mirrors the occurrence of a peak in $\varepsilon'(T)$ revealed in Fig. 2.

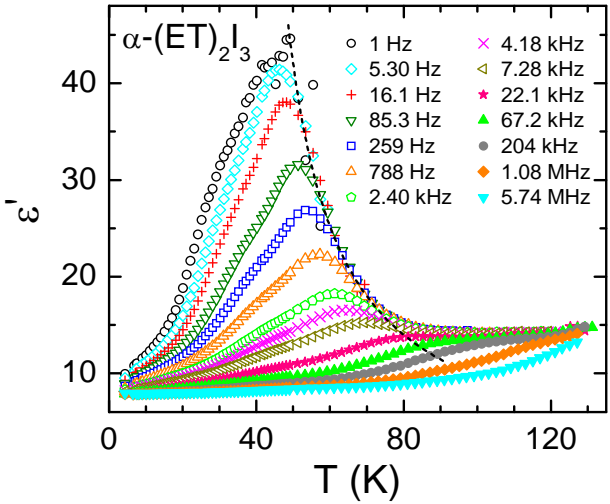

FIG. 2 (color online). Temperature-dependence of the dielectric constant for various frequencies. The dashed line was calculated assuming a Curie-Weiss law with $T_{CW} = 35$ K.

The right flank of the $\varepsilon'(T)$ peaks in Fig. 2 corresponds to the static dielectric constant $\varepsilon_s$ and can be formally described by a Curie-Weiss law with a characteristic temperature of $T_{CW} = 35$ K (dashed line). Although in relaxors deviations from Curie-Weiss behavior close to the peak are frequently found, this law at least provides a rough estimate of the quasistatic freezing temperature. $T_{CW}$ is of similar order as the Vogel-Fulcher-Tammann temperature of 29 K found from the above-mentioned fits of the $\varepsilon'(T)$ peaks, further corroborating the glass-like freezing of dipolar order in this temperature region. Of course, the Curie-Weiss formula cannot account for the additional non-ferroelectric relaxational behavior observed at $T > 75$ K in Fig. 2.

Figure 3 shows the frequency dependence of the real and imaginary parts of the permittivity, dielectric constant

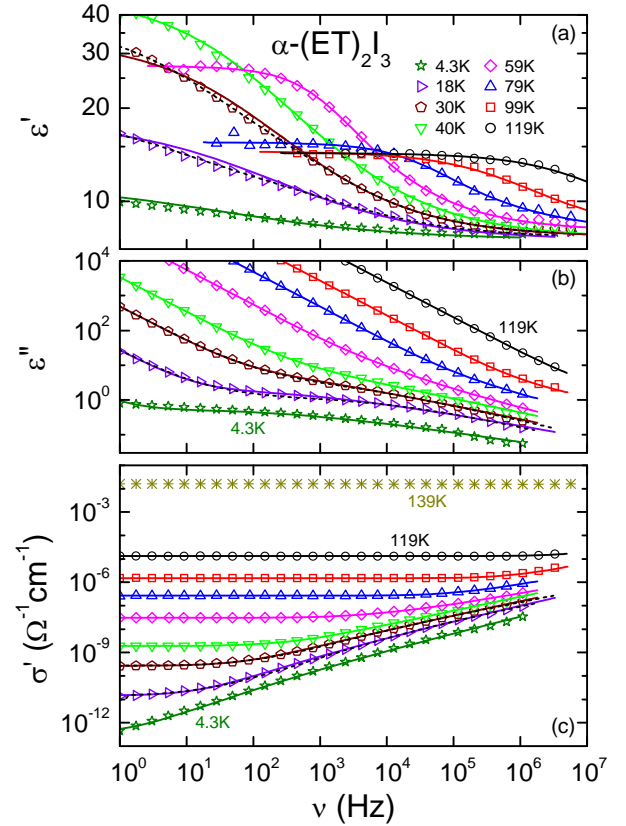

FIG. 3 (color online). Spectra of dielectric constant (a), loss (b), and real part of the conductivity (c), shown for various temperatures. The solid lines are fits with eq. (1). The dashed lines show alternative fits assuming the superposition of two relaxation processes, performed for the curves at 18 and 30 K.

For relaxational processes, the loss $\varepsilon''(\nu)$ should show a peak, located at a frequency corresponding to the point of inflection of the $\varepsilon'(\log \nu)$ curves. In Fig. 3(b), indications for such peaks are indeed found, especially for the lowest temperatures. However, they are strongly superimposed by a linear increase in the double-logarithmic representation of $\varepsilon''$ towards low frequencies, implying a $1/\nu$ behavior.

This can be ascribed to dc conductivity, which, via the general relation $\varepsilon'' = \sigma'/(2\pi\nu\varepsilon_0)$ ($\varepsilon_0$: permittivity of vacuum), leads to a $1/\nu$ divergence in the loss. Sometimes, $\varepsilon''$ data with strong charge-transport contributions are corrected for the dc conductivity using the relation $\varepsilon''_{corr} = \varepsilon'' - \sigma_{dc}/(2\pi\nu\varepsilon_0)$ to reveal the "pure" relaxation-caused loss peaks, which then are further analyzed [3,12,27]. However, one should note that some ambiguities in this correction procedure arise from the fact that the exact frequency position and shape of the peaks, obtained in this way, very critically depend on the accurate choice of the dc conductivity, which usually cannot be determined with sufficient precision (for a detailed discussion, see [13]).

Here we follow a less ambiguous approach by performing simultaneous fits of $\varepsilon'(\nu)$ and $\varepsilon''(\nu)$, *including* the contribution from dc charge transport. For the relaxational part, we used the empirical Cole-Cole function [41], commonly employed, e.g., to describe relaxations in glassy matter [33]. Thus, we finally arrive at:

$$\varepsilon' - i\varepsilon'' = \varepsilon_\infty + (\varepsilon_s - \varepsilon_\infty)/\left[1 + (i\omega\tau)^{1-\alpha}\right] - i\sigma_{dc}/(2\pi\nu\varepsilon_0) \qquad (1)$$

Here $\varepsilon_s$ is the static dielectric constant, $\varepsilon_\infty$ denotes the high-frequency limit of $\varepsilon'$, and $\alpha < 1$ controls the broadening of the loss peaks and corresponding $\varepsilon'(\nu)$ steps. The fit curves, shown by the lines in Fig. 3, provide a good description of the experimental spectra. Successful fits were also performed for spectra at additional temperatures, not included in Fig. 3 for clarity reasons. The resulting temperature dependences of the relaxation parameters are shown in the Supplemental Material [31].

A closer look at $\varepsilon'(\nu)$ in Fig. 3(a) reveals minor deviations between fits and experimental spectra at the lower temperatures, becoming most obvious for 18 and 30 K. Here the experimental relaxation steps are slightly more smeared out than the fit curves. Interestingly, this could not be taken into account by an adaption of the width parameter in the simultaneous fits of $\varepsilon'(\nu)$ and $\varepsilon''(\nu)$. A superposition of two separate relaxation processes can also lead to such a broadening. Obviously, the spectral features corresponding to these relaxations closely superimpose and an unequivocal deconvolution in the spectra is impossible. Nevertheless, we performed fits of $\varepsilon'(\nu)$ and $\varepsilon''(\nu)$ at 18 and 30 K using the sum of two Cole-Cole functions, which leads to perfect fits (dashed lines in Fig. 3). Based on this evaluation procedure, due to a correlation of parameters, no statement on their absolute values is possible. However, these fits at least demonstrate that the assumption of two relaxation processes is reasonable.

As revealed by Fig. 3(c) the conductivity spectra exhibit a plateau showing up at low frequencies, which arises from the dc conductivity. Except for the lowest temperature, where it is shifted out of the frequency window, the plateau is found for all frequencies and completely dominates the spectra at the two highest shown temperatures (note that 139 K is located above the CO transition where pure dc response is naturally expected). Comparing frames (b) and (c) of Fig. 3 reveals that the increase of $\sigma'(\nu)$ at high frequencies is completely governed by the relaxation process. (It should be again noted here that $\sigma'$ and $\varepsilon''$ are directly related via $\sigma' = \varepsilon'' \varepsilon_0 2\pi\nu$.) This leads to the crossover to weaker temperature dependence of $\sigma'(T)$ observed at low temperatures in Fig. 1. At high temperatures, the strongly increasing conductivity becomes so high that the relaxation contribution becomes submerged and no longer is visible in the dielectric loss. However in $\varepsilon'$ [Fig. 3(a)] it still can be detected because the dc conductivity only contributes to the imaginary part of the permittivity.

Interestingly, relaxor ferroelectricity, with Curie-Weiss behavior extending down to temperatures close to the peak in $\varepsilon'$, was also detected in the related charge-transfer salts $\kappa$-(ET)$_2$Cu$_2$(CN)$_3$ [5] and $\beta'$-(ET)$_2$ICl$_2$ [9]. Moreover, in Ref. [42], dielectric measurements of $\alpha$-(ET)$_2$I$_3$ for **E**||**c** performed in a limited temperature range, $T \leq 50$ K, revealed a decrease of $\varepsilon'(\nu)$ with frequency. While the absolute values of $\varepsilon'$ at low frequencies reported in [42] were higher than those of the present work, the observed frequency and temperature dependence is consistent with relaxational behavior and at least qualitatively agrees with $\varepsilon'(\nu)$ deduced from the present investigation as shown in Fig. 3(a). The present temperature-dependent measurements, performed for higher temperatures up to $T_{CO}$ (Fig. 2) reveal that in fact a frequency-dependent peak shows up in $\varepsilon'(T)$, resembling relaxor behavior.

### B. Polarization switching

To further check for ferroelectric order in $\alpha$-(ET)$_2$I$_3$, so-called "positive-up-negative-down" (PUND) measurements [43] were performed (Fig. 4). Here a sequence of trapezoid field pulses is applied to the sample (left inset). The current responses of the first and third pulse show peaks which occur when the electric field $|E|$ exceeds a threshold level of the order of 20 kV/cm. Obviously, here the field leads to a switching of the macroscopic polarization, generating a reorientation of the dipolar moments within the ferroelectric domains and, thus, a peak in $I(t)$. This notion is strongly supported by the absence of any significant polarization current at the second and forth pulse: Here the polarization was already switched by the preceding pulse and no further dipolar reorientation is expected. This finding corroborates the intrinsic nature of the polarization currents found for pulses I and III [43]. A closer inspection of Fig. 4 reveals that in the vicinity of the main polarization peaks a sequence of minor peaks is observed. By performing several PUND measurements on the same sample, we found that the sequence of peak events is of stochastic nature. Currently, we can only speculate about a connection of this finding to the suggested clusterlike nature of the electronic ferroelectricity in $\alpha$-(ET)$_2$I$_3$ and this interesting phenomenon certainly deserves further investigation. In addition to the polarization-induced peaks, the measured $I(t)$ is found to be approximately proportional to the applied trapezoid-shaped voltage pulses. This finding arises from charge transport within the sample. The flanks of the $I(t)$ trapezoids show a slight curvature, indicating nonlinear I-V characteristics as reported in [42,44].

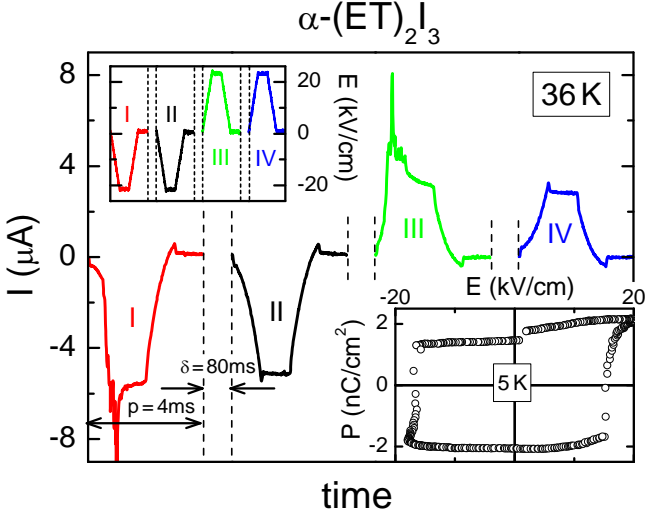

FIG. 4 (color online). Time-dependent current obtained from PUND measurements performed at 36 K with waiting time $\delta$ and pulse width $p$. Left inset: Excitation signal. Right inset: Polarization-field hysteresis curve at 5 K.

At the lowest temperatures investigated, the conductivity is sufficiently low to check for the occurrence of typical ferroelectric hysteresis curves seen in a conventional field-dependent polarization measurement. The right inset of Fig. 4 shows such a curve, providing further evidence for ferroelectric ordering in $\alpha$-(ET)$_2$I$_3$. The rather small absolute value of the saturation polarization of the order of 2 nC/cm$^2$ arises from the fact that the measurement temperature is far below the freezing temperature of about 35 K and only a small fraction of polar domains can be switched. In contrast, the polarization calculated from the PUND experiments at 36 K is much larger and reaches about 150 nC/cm$^2$ [31].

## IV. DISCUSSION

Overall, the polarization results of Fig. 4 provide strong evidence for ferroelectricity in $\alpha$-(ET)$_2$I$_3$. This is in accord with the absence of inversion symmetry of the crystal lattice indicated by SHG experiments [4,19]. Recently, ferroelectricity was also invoked to explain electric-field-pulse experiments on $\alpha$-(ET)$_2$I$_3$, assuming the simultaneous existence of charge-density waves and ferroelectric-like domains within the ET planes [45]. As $\alpha$-(ET)$_2$I$_3$ exhibits a combination of bond- and site-centered CO, as discussed above, electronic degrees of freedom causing the polar ordering in this material [1] seem the most likely. However, we stress that considering purely electronic effects is likely to be an oversimplification of the polarization process which may also include minor ionic displacements. For example, for the charge-transfer complex tetrathiafulvalene-p-chloranil a combination of weak ionic and strong electronic polarization was unequivocally proven [46,47].

In contrast to $\kappa$-Cl [7], the peak temperatures in $\varepsilon'(T)$ (Fig. 2) reveal a significant shift with frequency, reminding of the typical behavior of relaxor ferroelectrics [5,9,39,40]. Moreover, in $\kappa$-Cl, indications for ferroelectric ordering were found at the same temperature as a strong reduction of the conductivity, ascribed to CO [7,13] (see [11,12] for an alternative view). However, in $\alpha$-(ET)$_2$I$_3$ the signatures of ferroelectricity only show up far below $T_{CO} \approx 133$ K (Fig. 2). Thus, in contrast to $\kappa$-Cl, in $\alpha$-(ET)$_2$I$_3$ the CO transition and occurrence of ferroelectric order appear to be less closely coupled.

If assuming that the detected dielectric anomalies indeed imply relaxor ferroelectricity (see below for alternative explanations), we propose the following tentative picture to explain this finding: It seems reasonable to ascribe the decoupling of the CO transition and ferroelectric ordering to the particular structure of $\alpha$-(ET)$_2$I$_3$ where the one-dimensional molecular stacks I with polar order are separated by the non- or less-polar stacks II (cf. inset of Fig. 1). Consequently, the coupling between the polar stacks may be only weak, hampering the formation of three-dimensional long-range ferroelectric order. Therefore, at temperatures just below $T_{CO}$ no long-range ordering sets in and instead relaxational behavior is found arising from the relaxation of single dipoles or several dipoles that are ferroelectrically correlated within one chain. Single-dipole relaxation may well arise also in stacks II, lacking any dimerization. Only when the temperature is further lowered, three-dimensional ferroelectric order of the dipoles in stacks I finally forms, however, of cluster-like short-range nature only, leading to the signature of relaxor ferroelectricity (Fig. 2). Interestingly, when closely inspecting $\varepsilon'(T)$ in Fig. 2, small shoulders at the left flanks of the relaxation curves, especially at the lower frequencies (i.e. temperatures), are revealed that indicate the presence of an additional relaxation process (e.g., at about 35 K for the 259 Hz curve). As discussed above, an analysis of the frequency dependence of the permittivity (Fig. 3) also points to two relaxation processes. Thus it seems well possible that the relaxation of single, not ferroelectrically correlated dipoles on stack II still persists down to low temperatures.

As mentioned above, relaxor ferroelectricity was also reported for $\kappa$-(ET)$_2$Cu$_2$(CN)$_3$ [5] and $\beta'$-(ET)$_2$ICl$_2$ [9]. Within the scenario of electric-dipole-driven magnetism suggested in [7], the absence of long-range ferroelectric order in the first system is consistent with the occurrence of a spin-liquid state [48]. For $\beta'$-(ET)$_2$ICl$_2$, charge disproportionation within dimers and frustration between ferroelectric and antiferroelectric interactions due to spin-charge coupling was suggested to explain the dielectric behavior [9]. In the present nonmagnetic system [49], we propose that the peculiarities of the α structure of the ET planes, with alternating dimerized and undimerized stacks, prevent the formation of canonical long-range ferroelectric order. Alternatively, the competition of a fully ordered ferroelectric state and some other quasi-degenerate ground state could also explain the possible relaxor ferroelectricity in α-$(ET)_2I_3$. This may be in line with the simultaneous occurrence of a CDW and ferroelectric-like state proposed in Refs. [23,24,45]. In contrast to the recently found "charge-cluster glass" in θ-$(ET)_2RbZn(SCN)_4$, where long-range CO is suppressed due to geometrical frustration [50], in α-$(ET)_2I_3$ the CO is fully developed while the polar order shows glass-like characteristics.

It should be noted that space-charge effects can lead to non-intrinsic Maxwell-Wagner relaxations [25,26] and one may ask if the present finding of relaxational behavior in α-$(ET)_2I_3$ can be explained in this way. Space charges can be caused by electrode polarization, grain boundaries, or other heterogeneities. However, for the single-crystalline samples discussed here, where the high degree of crystalline perfection is reflected in a sharp CO transition, significant contributions from grain boundaries or other internal heterogeneities can be safely excluded. Therefore, surface-related effects, e.g., Schottky diodes arising from the metallic contacts, are the only candidates that could give rise to non-intrinsic relaxations. For any surface-related effect, marked differences of the dielectric constant are expected for samples with different area-to-thickness ratios because only the surface dominates the dielectric response while the dielectric constant is calculated from the measured capacitance $C$ using $\varepsilon' \propto C/(A/d)$ [26]. The area-to-thickness ratios of the two investigated crystals differ by more than a factor of two. Therefore the nearly identical results obtained for the two crystals [31] make it unlikely that there are any significant contributions from non-intrinsic surface effects.

Moreover, space-charge effects usually lead to conventional relaxational behavior with an only weakly temperature-dependent static dielectric constant, in marked contrast to the present finding of a strongly temperature-dependent $\varepsilon_s(T)$ [25,26]. A prominent example is $LuFe_2O_4$, for which electronic ferroelectricity was reported [51] but whose dielectric response later on was proven to be dominated by non-intrinsic space-charge effects [52]. Its $\varepsilon'(T)$ curves qualitatively differ from the present results. However, within the Maxwell-Wagner scenario strong anomalies in $\varepsilon'$, as the peak documented in Fig. 2, may arise when the intrinsic bulk conductivity exhibits a strong variation, too. An example is the artificial magneto-dielectric effect that is generated by a magnetic-field induced variation of the conductivity [53]. In the present case, indeed a conductivity anomaly is observed but it occurs at $T_{CO} \approx 133$ K while the $\varepsilon'$ peak shows up at about 40 - 60 K, where the conductivity exhibits a smooth variation only (Fig. 1). Thus, such a scenario does not apply for α-$(ET)_2I_3$.

It also should be noted that the conductivity jump at $T_{CO}$ (Fig. 1) proves that intrinsic behavior is detected. According to the Kramers-Kronig relation, within the Maxwell-Wagner scenario it is not possible that the measured conductivity is of intrinsic nature while the dielectric constant is purely non-intrinsic. As shown in Refs. 25 and 26, any contributions from Maxwell-Wagner effects vanish at low temperatures. If the dielectric response around $T_{CO}$ is intrinsic, this should also be the case for the results at all lower temperatures.

In summary, space-charge effects can always play a role in samples that are not completely insulating (but one should note that α-$(ET)_2I_3$ is rather insulating, reaching a dc conductivity of about $4\times10^{-9}$ $\Omega^{-1}cm^{-1}$ at the temperature of the $\varepsilon'$ peak). However, for the reasons discussed above, for α-$(ET)_2I_3$ a non-intrinsic space-charge origin of the observed relaxor like behavior at least seems very unlikely.

In Ref. [19], the observation of "180° polar domains growing in the ferroelectric phase" was reported based on optical SHG interferometry. The occurrence of these domains below the metal-insulator transition temperature and their rather large size (several 100 μm) seem to contradict the present indications of relaxor ferroelectricity (i.e., short-range clusterlike order) occurring rather far below $T_{CO}$. It should be noted that SHG detects the breaking of centric symmetry [20,21] and the findings of Ref. [19] prove the coherent arrangement of non-centric units, i.e. essentially structural domains. Of the 21 non-centrosymmetric crystallographic point groups only 10 are polar with a unique polar axis [22]. Moreover, materials belonging to these 10 polar groups are called ferroelectric only if their polarization is switchable [22]. Thus, dielectric methods are required to unequivocally prove ferroelectricity and to clearly detect the corresponding domains. We also want to remark that the wavelengths used in the optical experiments performed in Ref. [19] would prevent the detection of the polar nanodomains often assumed to exist in relaxor ferroelectrics [39,40]. If these considerations can resolve the apparent discrepancy between the results in Ref. [19] and the present study is not clear at present, and more experimental work is necessary to clarify this issue.

The present findings complement the report of ferroelectricity (and thus multiferroicity) in κ-Cl [7]: While for this material the occurrence of CO is controversial [11,12,13], it is well established in α-$(ET)_2I_3$. Just as proposed for κ-Cl [7], electronic dipolar degrees of freedom via CO lead to ferroelectric order in α-$(ET)_2I_3$. It only can be detected in **c** direction but remains undetectable by in-plane experiments due to the high in-plane conductivity and the charge-density-wave like effects that seem to dominate the dielectric response within the planes [23,24].

Finally, we want to mention that in a recent work, based on nonlinear I-V measurements, a Kosterlitz-Thouless-type transition was suggested to occur in α-$(ET)_2I_3$ at a temperature $T_{KT}$ of about 30 - 45 K [44]. Interestingly, this is just the region where the dielectric anomaly with $T_{CW} \approx 35$ K is observed in Fig. 2. In Ref. [44], for $T > T_{KT}$ thermal excitations of electron-hole pairs from the CO state were proposed to occur, which may well lead to a relaxational response in dielectric spectra. While the details are yet to be

## V. SUMMARY

In summary, via dielectric and polarization experiments, we found clear evidence for ferroelectric ordering in $\alpha$-$(ET)_2I_3$. We propose that it arises from electronic charge ordering on the dimerized ET molecules located in stacks I of the $\alpha$-phase structure. Moreover, we find indications for two relaxation processes: One we suggest to be caused by single dipoles or small clusters of dipoles located in stacks I and/or II. The second one resembles the typical relaxation behavior of a relaxor ferroelectric, caused by three-dimensionally ordered ferroelectric clusters, whose motion freezes in the region around 35 K. In contrast to ferroelectric $\kappa$-Cl, we tentatively propose that the formation of long-range electronic ferroelectricity in $\alpha$-$(ET)_2I_3$ is prevented by the peculiarities of the $\alpha$-phase structure of the ET planes.

It should be noted, however, that relaxor ferroelectricity at about 35 K apparently contradicts the common believe that $\alpha$-$(ET)_2I_3$ already becomes ferroelectric below $T_{CO} \approx 133$ K [4,19], which seems to be supported by the transition into the polar P1 structure as revealed by X-ray diffraction measurements [17]. We want to remark that this transition is of structural nature [17,29]. The ferroelectricity in $\alpha$-$(ET)_2I_3$, however, most likely is of predominantly electronic nature as in related systems, showing very similar relaxor-like dielectric behavior [5,6,8,9]. The discrepancy between the occurrence of a polar structure at $T_{CO}$ and our detection of relaxor ferroelectricity at about 35 K may well reflect the disparity of ionic and electronic polar degrees of freedom as also discussed for other systems [46,47]: The ionic ones form a polar structure at $T_{CO}$ while the electronic ones show clusterlike ferroelectric ordering at much lower temperatures. In any case, we do not claim here that we have definitely excluded ferroelectricity below $T_{CO} \approx 133$ K by our results. Other explanations of the observed relaxor behavior have to be considered, too. For example, the one-dimensional charge-ordered stacks or the zigzag chains within the ET planes [17,29] could exhibit defects that cause the complex electric response. In any case, the switchability of the polarization revealed by our investigations (Fig. 4) clearly proves the occurrence of ferroelectricity at low temperatures in $\alpha$-$(ET)_2I_3$. Its very interesting anomalous dielectric response at temperatures below $T_{CO}$ (Figs. 2 and 3), closely resembling that of the electronic relaxor ferroelectrics $\kappa$-$(ET)_2Cu_2(CN)_3$ [5] and $\beta'$-$(ET)_2ICl_2$ [9], calls for further investigations to finally clarify its microscopic origin

## ACKNOWLEDGMENTS

This work was supported by the Deutsche Forschungsgemeinschaft through the Transregional Collaborative Research Centers TRR 80 and TRR 49 and by the BMBF via ENREKON.